\documentclass{PoS}

\title{Parton correlation functions and factorization in deep inelastic scattering}

\ShortTitle{Parton correlation functions and factorization in DIS}

\author{\speaker{Ted C. Rogers}\\
        The Pennsylvania State University\\
        Email: \email{rogers@phys.psu.edu}}

\abstract{We outline the basic properties of a pertubative QCD factorization formalism 
that maintains exact over-all kinematics in both the initial and final states.  
Such a treatment requires the use of non-perturbative factors that depend on all 
components of parton four-momentum.  These objects are referred to as parton correlation functions.
We describe the complications faced in defining parton correlation functions and discuss recent progress.  Emphasis is placed on the need for 
precise operator definitions in a complete derivation of factorization.}

\FullConference{8th International Symposium on Radiative Corrections (RADCOR)\\
		 October 1-5 2007\\
		 Florence, Italy}

\begin{document}

\section{Exact over-all kinematics and the need for a generalized treatment of factorization}
 
The standard collinear factorization theorems of 
perturbative QCD (pQCD) rely on a number of kinematical 
approximations that change the momentum of the final 
state particles.  In inclusive deep inelastic scattering (DIS), 
for example, the 
struck parton is assumed to have a non-zero component of 
four-momentum only in the plus direction corresponding to the 
target beam direction. 

It has been known for some time that neglecting the
transverse component of the struck parton momentum is not always valid.
This has lead to studies of ``$k_t$-factorization'' involving $k_t$-unintegrated 
(or just unintegrated) parton distribution functions~\cite{smallx}.
More recently, it has been noted
that in certain cases over-all four-momentum conservation (involving both
transverse momentum and parton invariant energy) must be enforced 
to avoid making large errors, particularly when the details 
of final states are important~\cite{Collins:2005uv}.  
This has motivated the formulation of factorization theorems 
in which \emph{no} approximation is made on 
the momenta of initial and final states.
The non-perturbative objects in such a factorization formula will
depend on all components of parton four-momentum.
We call these fully unintegrated objects parton correlation functions (PCFs).
In these proceedings we outline the basic 
structure of the fully unintegrated formalism proposed in~\cite{CRS}.  
In addition, we re-emphasize the need for exact operator 
definitions for the non-perturbative factors in a factorization 
formalism.

\section{The role of operator-based definitions}

Having non-perturbative factors rooted in 
operator definitions is important for the derivation 
of a reliable factorization formula.  To understand this, let us briefly
review the basic requirements of a factorization theorem:
\begin{itemize}
\item For a given process with hard scale $Q$, a factorization formula exists if
the cross section
can be written approximately as a generalized 
product of several factors.  The hard scattering coefficient $C$ 
should involve only  
lines that are off-shell by order $Q$ and can be calculated 
explicitly in pQCD.  The other factors involve infrared and collinear lines and parameterize 
the non-perturbative physics.  Errors should be suppressed by powers of $\Lambda/Q$ where $\Lambda$ is
a typical hadronic mass scale.

\item The non-perturbative factors 
must be parameterized by experimental data. But if  
factorization formulae exist for different 
processes, and involve the \emph{same} 
non-perturbative factors, then one can 
parameterize a non-perturbative factor in one 
experiment and use it in another 
to make first-principle predictions.
If this can be shown, we say that the 
non-perturbative factors are universal.

\end{itemize}
Operator definitions are what allow for 
a comparison of the soft and collinear factors 
used in different processes. 
Hence, they are needed if one is to have confidence in the second
bulleted statement above.
As an example, consider the simple case of 
totally inclusive DIS.
The definition of the integrated parton distribution
function should arise naturally from the sequence of approximations
needed to factorize the hadronic tensor. 
The hadronic tensor written using standard notation is,
\begin{equation}
\label{eq:struct}
W^{\mu \nu}(P,q) = 4 \pi^{3} \sum_{X} \langle p \left| J^{\mu} (0) 
\right| X \rangle \langle X \left| J^{\nu}(0) \right| p \rangle \delta^{(4)} (p_X - (p+q)). 
\end{equation}
Starting with this very basic formula and applying
certain approximations, one can derive the formula,
\begin{equation}
\label{eq:fact}
P_{\mu \nu} W^{\mu \nu}(P,q) = P_{\mu \nu} H^{\mu \nu}_j (q,\xi,\mu) \otimes f_{j/h}(\xi, \mu) + \mathcal{O}(\Lambda/Q \left| P_{\mu \nu} W^{\mu \nu}(P,q) \right| ),
\end{equation}
where $P_{\mu \nu}$ is defined to act on the electromagnetic vertices to 
project out a particular structure function, say $F_1$.
The function $f_{j/h}(\xi, \mu)$ is the usual expression for 
the fully integrated parton distribution function (PDF),
\begin{equation}
\label{eq:PDF}
f_{j/h} (\xi, \mu) = \int \frac{dw^-}{4 \pi} 
e^{-i \xi p^{+} w^-}  
\\\times
\langle p | \bar{\psi}(0,w^-,{\bf 0}_{T}) V^{\dagger}_{w}(u_{\rm J})
\gamma^{+} V^{}_{0}(u_{\rm J}) \psi(0) | p \rangle.
\end{equation}
Here $\psi$ is the quark field operator, and $V_{0}(u_J)$ is a Wilson line
operator in the direction $u_J = (0,1,{\bf 0}_t)$ needed to make the definition 
exactly gauge invariant.
The lowest order hard scattering matrix element is the 
usual one involving only the electromagnetic vertex. 
Using it in Eq.~(\ref{eq:fact}) reproduces the parton model.
Higher order corrections are calculated 
by considering more complex graphs and applying a sequence of 
subtractions to remove double counting.
The unregulated PDFs contains the usual 
ultraviolet divergences which are effectively removed 
using standard renormalization techniques, with a renormalization scale $\mu$.  
The resulting evolution equations describe 
the well-known scaling violations of DIS.  
The fact that the PDF $f_{j/h} (\xi, \mu)$ also appears
in the factorization formula for other processes means that it can be 
parameterized and used for future predictions.

One may hope to have the same powerful structure in a more general unintegrated 
formalism.
In a treatment that includes the transverse momentum of the struck parton, it is tempting to 
propose a differential 
definition for the $k_t$-unintegrated parton distribution,
\begin{equation}
\label{eq:uPDF1}
\mathcal{F}_{j/p}({\bf k}_t,\xi) \stackrel{??}{=} \left. \frac{\partial f_{j/h}(\xi, \mu^2)}{\partial \ln \mu^2} \right|_{\mu^2 = k_T^2}.
\end{equation}
Studies of small-x physics~\cite{smallx} suggest that this 
is probably correct to a good approximation in the small-$x$ limit.  
However, it is unclear in general whether there is a reliable sequence of 
approximations that allow (\ref{eq:uPDF1}) 
to be factorized out of the hadronic tensor with only a hard scattering coefficient left over.
The universality of the $k_t$-dependent PDF is thus called into question.

The situation with $k_t$-unintegrated PDFs is further complicated by the 
fact that the most obvious candidate for a definition is
unsuitable for use in factorization.
Namely, if we leave one of the integrals in Eq.~(\ref{eq:PDF}) undone, we obtain the seemingly natural definition,
\begin{equation}
\label{pdf1}
  \mathcal{F}_{j/p}(x,{\bf k}_{T},\mu)  
  \stackrel{??}{=} \int \frac{dw^- d{\bf w}_{T}}{16 \pi^{3}}
   e^{-i x p^{+} w^- + i {\bf k}_T \cdot {\bf w}_T}
\\\times 
 \langle p | \bar{\psi}_j(0,w^-,{\bf w}_{T}) V^{\dagger}_{w}(n) 
     \gamma^{+} V_{0} (n) \psi_j(0) | p \rangle.
\end{equation}
However, this definition acquires divergences from 
gluons moving with infinite rapidity in the outgoing quark direction making it
inappropriate for use as a PDF.  (See~\cite{Collins:2003fm} for a more detailed discussion.)
An additional problem, pointed out by Belitsky et al.~\cite{Belitsky:2002sm}, is that exact gauge 
invariance requires the insertion of a gauge link at light-cone 
infinity.  Recent work in providing consistent operator definitions for 
$k_t$-unintegrated PDFs has been done by Hautmann and Soper~\cite{Hautmann:2007cx}.

\section{A fully unintegrated formalism}

As already mentioned, there are situations where we 
need a formalism based on non-perturbative objects that 
depend on \emph{all} components of parton four-momentum.
An approach to factorization using PCFs was formulated in~\cite{CZ} 
for the case of a scalar field theory, and extended to 
the case of a gauge theory in~\cite{CRS}.\footnote{The derivation is 
so far only complete for the case of an abelian gauge theory}.  
We summarize these results now.  To save space we do not list
the actual definitions, but rather outline the basic structure
of their derivation.  For all details, the reader is referred to~\cite{CRS}.

\begin{figure*}
\centering
  \begin{tabular}{c@{\hspace*{5mm}}c}
    \includegraphics[scale=0.3]{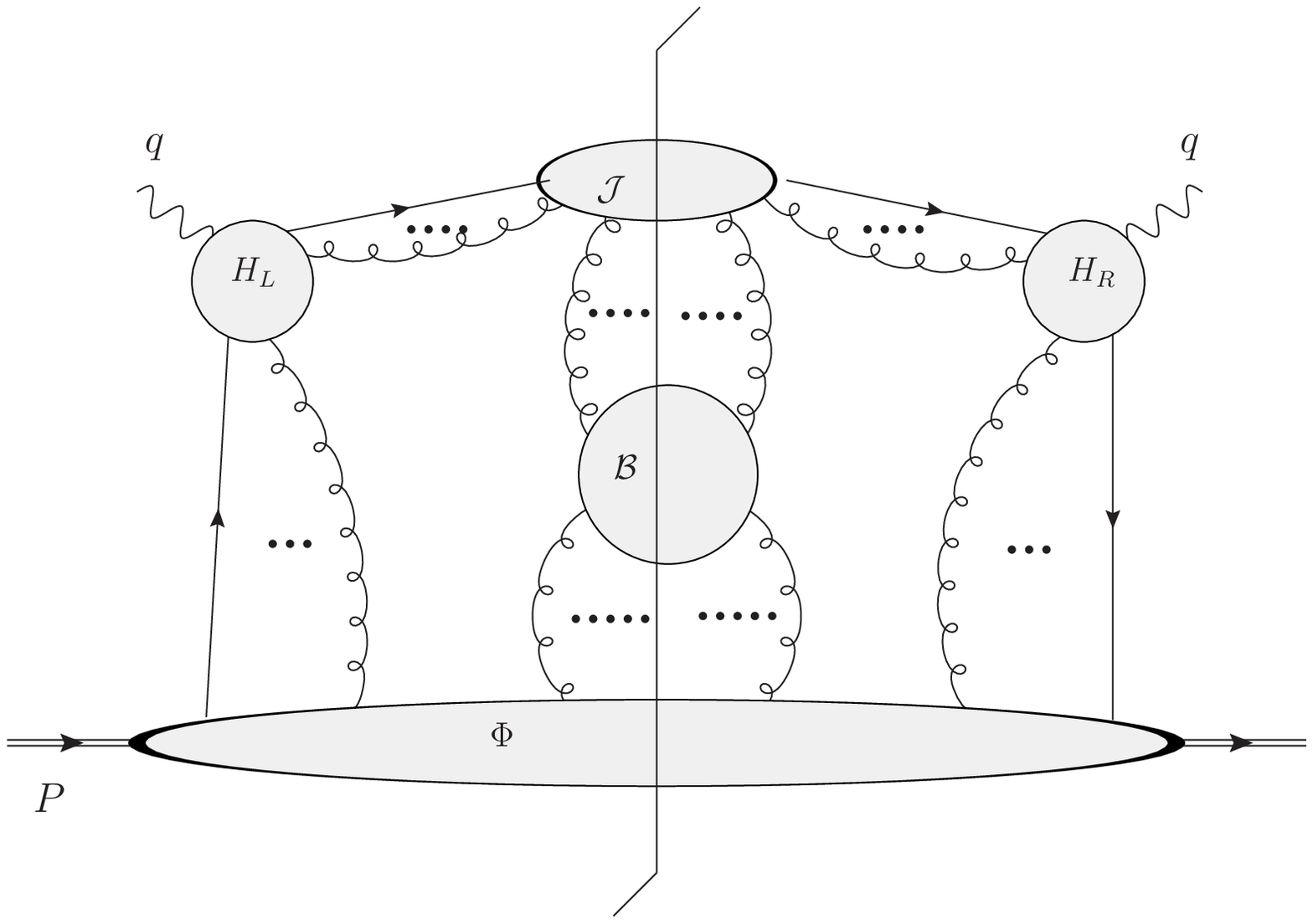}
    &
   \includegraphics[scale=0.3]{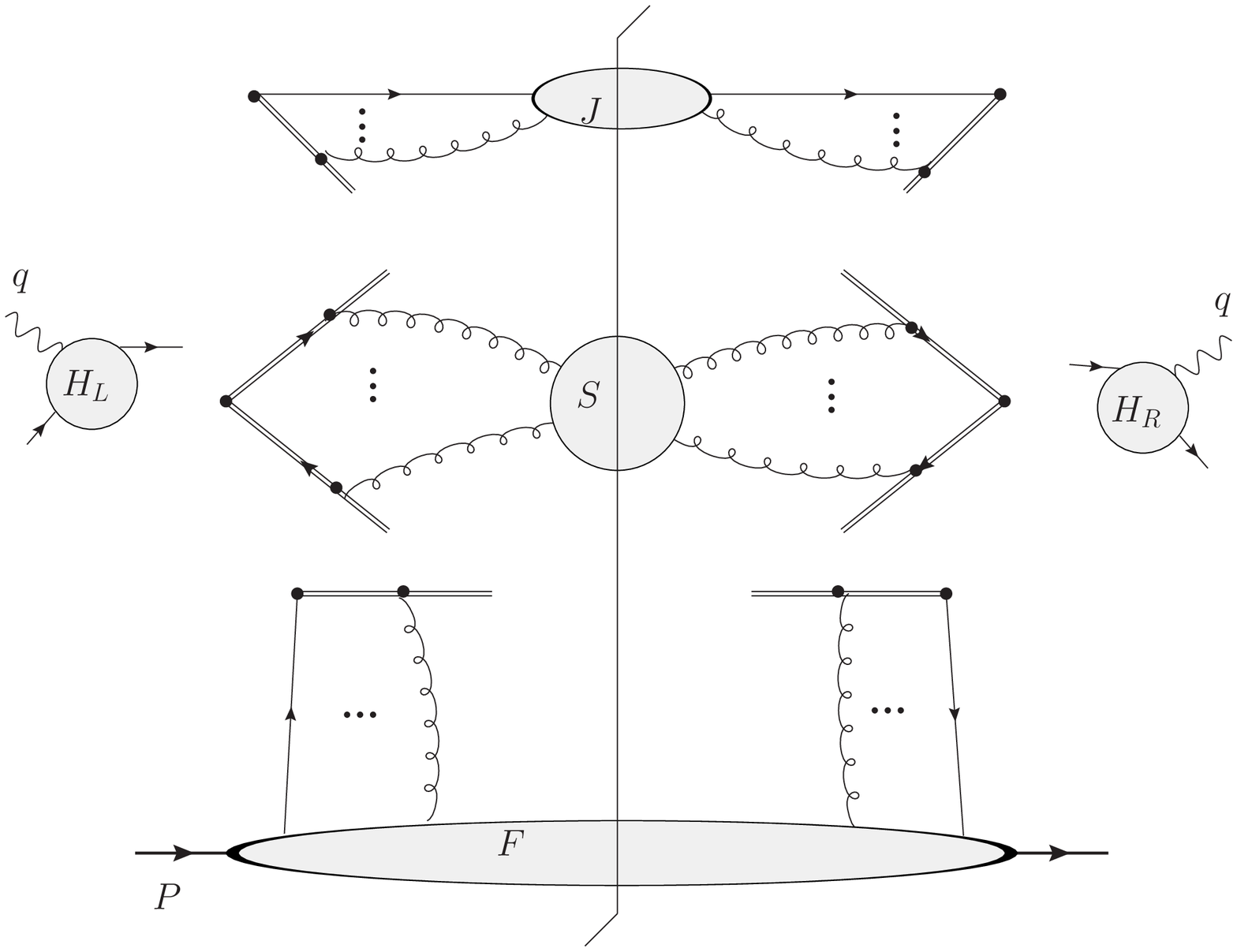}\\
(a) & (b)
  \end{tabular}
\caption{(a) General contribution to DIS at lowest order and (b) Structure of the 
graph after factorization.}
\label{LOdiags}
\end{figure*}

To avoid making errors of the type discussed in the first section,
one must begin with graphs of the general structure shown in Fig.~\ref{LOdiags}(a)
rather than the usual handbag diagram.  The bubbles represent 
sums of diagrams contributing to initial and final states.  
The extra gluons shown attaching the collinear bubbles to 
the hard scattering bubble represent possible extra target 
and final state collinear gluons.  In addition, there may be arbitrarily 
many soft gluons connecting the outgoing jets via a soft bubble.  
Before factorization, these bubbles are $\Phi$ for the target collinear 
lines, $\mathcal{J}$ for the jet collinear lines, and $\mathcal{B}$ for 
the soft lines.

It follows from general arguments~\cite{LS} that graphs with the topology 
of Fig.~\ref{LOdiags}(a) are the most general contributions 
to DIS involving a single outgoing jet.
Applying Ward identities to the sum of graphs 
with this allows the extra 
soft and collinear gluons to be disentangled into separate factors.
After topological factorization is achieved, the graph takes the
form shown in Fig.~\ref{LOdiags}(b).
The different PCFs include a soft factor $S$, a jet factor $J$, and 
a target factor(fully unintegrated PDF) $F$.  The PCFs are represented graphically
by the bubbles.  The double lines associated with each bubble are eikonal lines that correspond 
to Wilson lines in the definitions of the PCFs.
To be consistent with factorization, the definitions of the PCFs also
require double counting subtractions.  
(To avoid clutter, 
the double counting subtractions aren't shown explicitly in the figure.)
The factors S,J, and F can ultimately 
be shown to arise from operator definitions of the PCFs.
Schematically, the final factorization formula is,
\begin{equation}
\sigma = C \otimes F \otimes J \otimes S
+ \mathcal{O} \left( \left( \Lambda/Q \right)^{a} |\sigma | \right),
\end{equation}
where $\sigma$ is a measurable quantity such as a cross section 
or structure function and $a > 0$.

\section{Outlook}
  
Our  factorization formula is now more complicated than 
the usual one in Eq.~(\ref{eq:fact}).  Each PCF depends on several
parameters and each needs to be fitted to experimental data.  
For the fully unintegrated formalism to be practical, factorization formulae using the same PCFs will need
to be derived for a number of non-trivial processes.  
In addition to evolution in $\mu$, the evolution equations for the PCFs will
also involve evolution in other rapidity variables which act as effective cutoffs
on rapidity divergences.  One hope is to relate this type of evolution  
to more common approaches such as the CCFM equation.

As we have mentioned, the factorization formula represented schematically
in Fig.~\ref{LOdiags}(b) is only complete at lowest order in the hard scattering coefficient 
because it involves only one outgoing jet line.
However, this result is already quite useful because it now allows higher order corrections
to be obtained via double counting subtractions applied to more complicated graphs.
As in~\cite{CZ} for the scalar field theory, the hard scattering coefficients are
expected to be ordinary functions as opposed to the generalized functions (e.g. $\delta$-functions)
that appear in the hard coefficients of the standard integrated formalism.

\section*{Acknowledgments}
This talk is a summary of work done in collaboration with  John Collins and Anna Sta\'sto.
I would like to thank the organizers of RADCOR 2007 for their hospitality at this very productive
symposium.  This work was supported by the U.S. D.O.E. under grant number DE-FG02-90ER-40577.

\end{document}